\documentclass[twocolumn,tightenlines,aps,preprintnumbers,superscriptaddress,showpacs,floatfix]{revtex4}

\usepackage{graphicx}
\usepackage{epsf}
\usepackage{amssymb}
\usepackage{amsmath,amssymb,textcomp,array}

\newcommand{\half}{\mbox{\small $\frac{1}{2}$}}

\begin{document}

\preprint{LA-UR-07-0056}

\title{Cosmic Calibration: Constraints from the Matter Power
Spectrum and the Cosmic Microwave Background}
\author{Salman Habib}
\affiliation{T-8, MS B285, Los Alamos National Laboratory, Los
Alamos, NM 87545}
\author{Katrin Heitmann}
\affiliation{ISR-1, MS D466, Los Alamos National Laboratory, Los
Alamos, NM 87545}
\author{David Higdon}
\affiliation{CCS-6, MS F600, Los
Alamos National Laboratory, Los Alamos, NM 87545}
\author{Charles Nakhleh}
\affiliation{X-2, MS T087, Los Alamos National Laboratory, Los
Alamos, NM 87545}
\author{Brian Williams}
\affiliation{CCS-6, MS F600, Los
Alamos National Laboratory, Los Alamos, NM 87545}

\date{\today}

\begin{abstract}
  Several cosmological measurements have attained significant levels
  of maturity and accuracy over the last decade. Continuing this
  trend, future observations promise measurements of the statistics of
  the cosmic mass distribution at an accuracy level of one percent out
  to spatial scales with $k\sim 10h$Mpc$^{-1}$ and even smaller,
  entering highly nonlinear regimes of gravitational instability. In
  order to interpret these observations and extract useful
  cosmological information from them, such as the equation of state of
  dark energy, very costly high precision, multi-physics simulations
  must be performed. We have recently implemented a new statistical
  framework with the aim of obtaining accurate parameter constraints
  from combining observations with a limited number of simulations.
  The key idea is the replacement of the full simulator by a fast
  emulator with controlled error bounds. In this paper, we provide a
  detailed description of the methodology and extend the framework to
  include joint analysis of cosmic microwave background and large
  scale structure measurements. Our framework is especially
  well-suited for upcoming large scale structure probes of dark energy
  such as baryon acoustic oscillations and, especially, weak lensing,
  where percent level accuracy on nonlinear scales is needed.
\end{abstract}


\maketitle

\section{Introduction}

Over the last three decades observational cosmology has made
extraordinary progress in determining the make-up of the Universe and
its expansion history. The first precision observations were obtained
from the cosmic microwave background (CMB), beginning with the all-sky
temperature anisotropy measurements from COBE~\cite{smoot92} which
provided an encouraging confirmation of current theories of the early
Universe and the formation of large scale structure. Follow-up
measurements from the ground~\cite{cbi, dasi, acbar},
balloons~\cite{boomerang, archeops}, and space~\cite{spergel} have
resulted in constraints on the main cosmological parameters at better
than the $10\%$ percent level of accuracy, and the Planck satellite
mission~\cite{planck} promises even further improvement. But CMB
measurements are not the only observational source for precision
cosmology. Structure formation probes such as large-scale surveys of
the distribution of galaxies, and weak lensing and cluster surveys,
are reaching similar levels of statistical and systematic control, as
are supernova observations. These newer techniques, exemplified by
surveys such as the Sloan Digital Sky Survey (SDSS)~\cite{gunn06}
yield complementary data to the CMB to help determine the large-scale
description of the Universe~\cite{tegmark06}.

Precision measurements from several different cosmological probes have
revealed a highly unexpected result: roughly 70\% of the Universe is
made up of a mysterious dark energy which is responsible for a recent
epoch of accelerated expansion. Understanding the nature of dark
energy is the foremost challenge in cosmology today. Ground-based
telescopes and satellite missions have been proposed or are under
development to measure the equation of state parameter of dark energy
$w\equiv p/\rho$ ($p$ is the pressure and $\rho$ the density) at the
one percent level, and its time derivative to 10\%. Cosmic
structure-based methods to understand the nature of dark energy
include: baryon acoustic oscillations as probed by the large-scale
distribution of galaxies~\cite{bao}, weak lensing measurements of the
dark matter distribution~\cite{wl}, and measurements of the abundance
of clusters of galaxies~\cite{clusters}. All three probes require the
understanding of nonlinear physics at different length scales. At
small scales, in addition to gravity, baryonic physics plays an
important role, significantly complicating the modeling task.

As cosmological observations continue to improve, increasing demands
are placed on the underlying theory. Since cosmology is an
observational science, the role of theory in interpreting observations
is crucial to the success of the entire enterprise. Therefore, in
order to interpret and optimally design future observations,
theoretical predictions have to be at least as accurate -- preferably
more accurate -- than the observations. In different arenas of
cosmology, however, the individual levels of theoretical control are
far from uniform.

The growth and formation of large scale structure in the Universe
results from the action of the gravitational instability on primordial
fluctuations. Currently, by far the most favored scenario for
generating these fluctuations is perturbations from
inflation~\cite{inflation}, and theoretical predictions for most
simple inflationary models can be computed rather
precisely~\cite{uniform}, certainly better than the level of accuracy
set by near-future CMB observations. A key theoretical task lies in
connecting the primordial fluctuations to present-day observations.

Of all cosmological probes, our understanding of the physics of the
CMB is by far the most advanced. Because linear theory is applicable
in this case, observables such as the temperature power spectrum can
be determined with high accuracy, the linearized Einstein equations
and relevant Boltzmann equations being treated more or less fully as,
e.g., in Ref.~\cite{efst}. This approach is rather expensive, however,
and the development of more efficient methods has been an important
research activity in the last decade.  The most popular approach is
the line-of-sight integration method, the underlying algorithm in
codes like CMBFAST~\cite{cmbfast} or CAMB~\cite{camb} which are the
major resources used for CMB analysis today. A single run with such a
code takes only a few seconds. Since cosmological parameter estimation
can often involve tens of thousands of Markov Chain Monte Carlo (MCMC)
simulation runs, several approaches have been developed to replace
even these codes by some type of ``look-up'' technique. These methods
include purely analytic fits~\cite{teg00,Jim04}, combinations of
analytic and semi-analytic fits~\cite{kap02}, and interpolation
schemes based on a large set of training
runs~\cite{fendt06,auld06}. Most of the approximations are accurate at
the 5\% level over their range of validity, some being accurate at the
sub-percent level over a limited range of parametric variation. The
accuracy of all these schemes deteriorates very rapidly, however, if
the parameter ranges under consideration are expanded. Also, in
general it is non-trivial to extend these schemes to include
additional parameters or different data sets. We will return to some
of these issues in more detail in Appendix~\ref{appa} where we compare
these methods to those explained in this paper.

For large scale structure probes of cosmology the situation is very
different. The treatment of nonlinear physics can (mostly) no longer
be avoided, and depending on the scales of interest, baryonic physics
has to be treated accurately as well. In order to predict the matter
power spectrum or the halo mass function in the regimes of interest,
large, costly $N$-body codes have to be resorted to. Fits to the
matter power spectrum such as those given in Refs.~\cite{smith03,PD}
are accurate at the 10\% level, an order of magnitude shy of that
eventually required.

In the case of baryon acoustic oscillations, the relevant length
scales of interest are around 100$h^{-1}$Mpc, and it is sufficient to
carry out dark matter-only simulations (the understanding of
systematics with regard to galaxy properties may require incorporation
of extra physics). For upcoming, near-future weak lensing measurements
the scales of interest are around 10$h^{-1}$Mpc, once again, dark
matter only simulations being able to faithfully capture all the
physics relevant on those scales. Future ground-based
telescopes~\cite{lsst, panstarrs} and space missions, such as the
Joint Dark Energy Mission~\cite{jdem}, will push the weak lensing
scales beyond 1$h^{-1}$Mpc. At these scales baryonic physics affects
the matter power spectrum at the 10\% level~\cite{jing06,rudd} and
must be included in the simulations. Clusters of galaxies probe even
smaller scales -- in this case, an accurate treatment of gas physics
and astrophysical feedback mechanisms is absolutely
essential~\cite{clusters}.

All of these simulation tasks are major undertakings, even for the
``easiest'' cases, where pure dark matter simulations are sufficient,
i.e., baryon acoustic oscillations and weak lensing on scales
$>10h^{-1}$Mpc. A typical high-accuracy simulation for such studies
would be in the billion particle, Gigaparsec cubed volume class.  Such
a simulation carried out with the treecode HOT~\cite{warren}, one of
the world's most efficient $N$-body codes, requires roughly
30,000~Cpu-hours (compared to a few seconds of a CAMB run for CMB
predictions) on current hardware.  It is, therefore, immediately
obvious that -- at least in the foreseeable future -- a brute force
approach to running a large number of cosmological models with an
$N$-body code will not be feasible. One can envision running hundreds
of state-of-the-art simulations, but a number like tens of thousands
would be well out of reach.

It is therefore important to investigate how to extract robust
predictions for untried cosmological (and modeling) parameter settings
based on a relatively small number of very accurate simulations. A
successful framework for achieving these aims should:
\begin{enumerate}
\item require only a tractable number of costly simulations to create
  an accurate emulator which can replace the full simulator,
\item provide an optimal sampling strategy for the simulation runs to
  obtain the best possible performance of the emulator,
\item integrate the uncertainties from the emulator predictions into
  the parameter constraints,
\item be easy to extend to include diverse data sets,
\item be capable of handling a large set of cosmological parameters
  without catastrophically increasing the computational overhead.
\end{enumerate}
 
We have recently introduced such a statistical framework~\cite{HHNH}
in order to determine cosmological and model parameters and associated
uncertainties from simulations and observational data (for an overview
of the basic ideas see, e.g., Refs.~\cite{KOH} and \cite{GoldRou}). The
framework integrates a set of interlocking procedures: (i) {\em simulation
design} -- the determination of the parameter settings at which to
carry out the simulations; (ii) {\em emulation} -- given simulation output
at the input parameter settings, how to estimate the output at new,
untried settings; (iii) {\em uncertainty and sensitivity analysis} --
determining the variations in simulation output due to uncertainty or
changes in the input parameters; (iv) {\em calibration} -- combining
observations (with known errors) and simulations to estimate parameter
values consistent with the observations, including the associated
uncertainty. The last step enables predictions of new cosmological
results with a set of uncertainty bounds.

Our initial results are very promising. We find that only a relatively
small number of (sufficiently accurate) simulations appear to be
required in order to observationally constrain several cosmological
parameters at the few percent level. Partly, this is due to the fact
that in cosmology the response surface being modeled by the emulator
is very smooth, and partly due to the relatively narrow range of
variation for the prior values of some parameters. Our choices of
Gaussian process (GP) modeling for the emulator and the associated
stratified sampling procedure, described in Section~\ref{sec:emulator}
below, are known to function particularly well under these
circumstances.

In this paper we discuss the methodology behind the framework in some
detail. For concreteness we focus on a simple example application:
Estimation of five cosmological parameters from dark matter structure
formation simulations and a synthetic set of ``WMAP + SDSS''
measurements of the matter power spectrum. The statistical framework
is introduced in Section~\ref{sec:stat}, along with an explanation of
how the synthetic datasets were constructed, the results of various
tests, as well as the final results from the estimated posterior
distribution of the five cosmological parameters. The framework is
extended in Section~\ref{sec:combined} to combine disparate
measurements, in the present case, separate measurements of the matter
power spectrum and of the CMB temperature anisotropy. Conclusions and
future directions are presented and discussed in
Section~\ref{sec:dis}. In Appendix~\ref{appa} we compare our approach
for fast calculation of the CMB temperature anisotropy with other
interpolation schemes.

\section{The Statistical Framework}
\label{sec:stat}
In this section we describe our statistical methodology aimed at
confronting physical observations with output from a simulation model
in order to best infer unknown model parameters. As a specific
application example, we consider a single set of synthetic
observations $y$ of the mass power spectrum (Fig.~\ref{plotone}) along
with a finite sample set of mass power spectra derived from $N$-body
simulations run with different choices of cosmological parameters.

The simulation model requires $p_\theta$-vector $\theta^*$ of input
parameter settings (in our case cosmological parameters) in order to
produce a mass power spectrum $\eta(k;\theta^*)$ ($k$ being the
wavenumber).  The simplest possible assumption is to postulate that the
vector of observations $y$ is a noisy version of the simulated
spectrum $\eta(k;\theta)$ at the true setting $\theta$:
\begin{equation}
 y = \eta(k;\theta) + \epsilon,
\end{equation}
where the error vector $\epsilon$ is normal, with zero mean and variance
$\Sigma_y$.  Given a prior distribution $\pi(\theta)$ for the true
parameter vector $\theta$, the resulting posterior distribution
$\pi(\theta|y)$ for $\theta$ is given by
\begin{equation}
 \pi(\theta|y) \propto L(y|\eta(k;\theta)) \cdot \pi(\theta),
\end{equation}
where $L(y|\eta(k;\theta))$ comes from the normal sampling model for the
data 
\begin{equation}
L(y|\eta(k;\theta)) = \exp\left\{ -\frac{1}{2} (y-\eta(k;\theta))^T
 \Sigma_y^{-1} (y-\eta(k;\theta)) \right\}.  
\end{equation}
In principle, this posterior distribution could be explored via Markov
Chain Monte Carlo (MCMC) as has become a standard practice in
cosmological data analysis. However, if a single evaluation of
$\eta(k;\theta)$ requires hours (or days) of computation, a direct
MCMC-based approach is infeasible.
 
Our approach deals with this computational bottleneck by treating
$\eta(\cdot)$ as an unknown function to be estimated from a fixed
collection of simulations
$\eta(k;\theta^*_1),\ldots,\eta(k;\theta^*_m)$.  This approach
requires a prior distribution for the unknown function $\eta(\cdot)$,
and treats the simulation output $\eta^* = (\eta(k;\theta_1^*),
\ldots,\eta(k;\theta^*_m))^T$ as additional data to be conditioned on
for the analysis.  Hence there is an additional component of the
likelihood obtained from the sampling model for $\eta^*$ given the
underlying function $\eta(\cdot)$ which we denote by
$L(\eta^*|\eta(\cdot))$.
 
For this case, the resulting posterior distribution has the general
form
\begin{equation}
\pi(\theta,\eta(\cdot)|y,\eta^*) \propto L(y|\eta(\theta))\cdot
L(\eta^*|\eta(\cdot)) \cdot \pi(\eta(\cdot)) \cdot \pi(\theta),
\label{eq:post0}
\end{equation} 
which has traded direct evaluations of the simulator model for a more
complicated form which depends strongly on the prior model for the
function $\eta(\cdot)$.  Note that the marginal distribution for
$\theta$ will be affected by uncertainty regarding $\eta(\cdot)$.

In the following subsections, we describe in detail a particular
formulation of Eqn.~(\ref{eq:post0}) in the context of the synthetic
mass power spectrum application which was earlier used in
Ref.~\cite{HHNH}.  This formulation has proven fruitful in a variety
of physics and engineering applications which combine field
observations with detailed simulation models for inference.  In
particular we cover approaches for choosing the $m$ parameter settings
at which to run the simulation model, and a (prior) model -- or
emulator -- which describes how $\eta(\cdot)$ is modeled at untried
parameter settings.  Section~\ref{sec:full} describes how the observed
data is combined with the simulations and the emulator to yield the
posterior distribution.  In the following section we will demonstrate
how this formulation can be extended to combine information from
different data sets from galaxy surveys and cosmic microwave
background measurements.

\subsection{The Synthetic Power Spectrum}
\label{sec:data}

A synthetic data set has the key advantage that the underlying
cosmological parameters are known {\em a priori}. This allows direct
testing of any proposed statistical procedure for estimating these
parameters. In order to generate the synthetic power spectrum we run
ten realizations of a given cosmology in a large cosmological volume
with the particle mesh (PM) code MC$^2$ (for more information and
comparison results against other codes, see Ref.~\cite{HRWH}). The
matter transfer function used to set the initial conditions is
generated using CMBFAST~\cite{cmbfast}. Averaging over the ten
realizations (each of which covers a simulation volume of
450$h^{-1}$Mpc cubed) produces a smooth power spectrum suppressing
uncertainties due to cosmic variance.

\begin{figure}[t]
\begin{center}
\includegraphics[width=80mm,angle=-90]{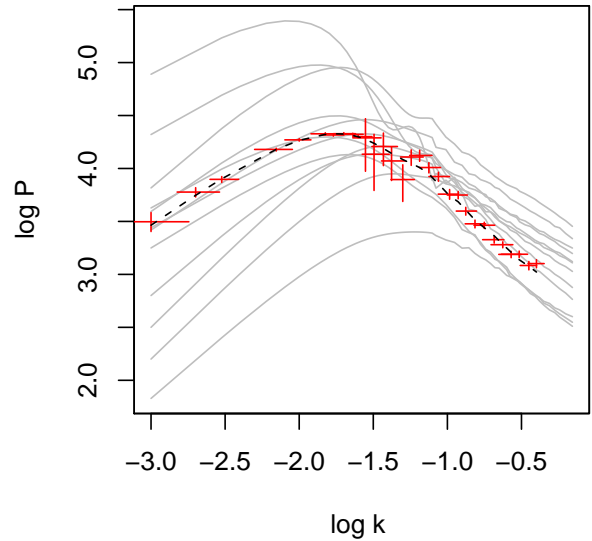}
\end{center}
\caption{Twenty-eight data points mimicking a combined data set from a
  large scale structure survey and CMB data. The error bars are chosen
  to match the power spectrum calculation in Ref.~\cite{tegmark}. The
  dashed line shows the original input power spectrum from ten
  realizations, while the gray lines show a subset of the 128
  simulated power spectra, discussed in Section~\ref{design}.}
\label{plotone}
\end{figure}

We consider five input parameters for the power spectrum in a
$\Lambda$CDM cosmology, the spectral index, $n$, the Hubble parameter
in units of 100~km/s/Mpc, $h$, the normalization of the amplitude as
specified by $\sigma_8$, and the dark matter and baryonic
contributions to the matter density specified as fractions of the
critical density, $\Omega_{\rm CDM}$ and $\Omega_{\rm b}$,
respectively
\begin{equation}
\theta=(n,h,\sigma_8,\Omega_{\rm CDM}, \Omega_{\rm b}).
\end{equation}
The further assumption of spatial flatness, $\Omega_{\rm tot}=1$, then
uniquely fixes the contribution of the cosmological constant,
$\Omega_{\Lambda}$. The five input parameters chosen for generating
the synthetic power spectrum are
\begin{equation}
\label{cosm}
\theta=(0.99,0.71,0.84,0.27,0.044).
\end{equation}
We match the nonlinear power spectrum at $k=0.1h$Mpc$^{-1}$ to the
linear power spectrum in order to increase the $k$-range down to very
large length scales, $k=0.001h$Mpc$^{-1}$. Next, we choose 28 points
from this combined power spectrum spaced roughly in the same bins as a
typical matter power spectrum extracted from combined CMB and
large-scale structure observations, more specifically, those for
WMAP~\cite{spergel} in the low $k$-range transitioning to values
typical for SDSS data~\cite{sdss} in the large $k$-range. Finally, the
points are moved off the base power spectrum according to a Gaussian
distribution with 1-$\sigma$ confidence. The resulting ``measurement''
and the smooth input power spectra are shown in
Fig.~\ref{plotone}. These synthetic observations will be the
underlying data set for verifying and testing the statistical analysis 
framework which we now discuss in detail.

\subsection{The Simulation Design}
\label{design}

Ongoing and near-future observations set stringent requirements on the
accuracy of theoretical predictions for observables such as the power
spectrum and the halo mass function. It will soon be insufficient to
use analytic fits for the power spectrum (see, e.g.,
Refs.~\cite{PD,smith03}) or Press-Schechter like fits (e.g.,
Refs.~\cite{PS, ST,WAHT,HLHR,Reed06,Lukic07}) for the mass
function. Fully nonlinear treatments based on simulations will be
needed, especially if the aim is to get reliable results on small
scales ($k \ge 0.2h^{-1}$Mpc).  As discussed earlier, such simulations
can be very costly, especially if they are not restricted only to dark
matter, but also include gas physics. In addition, the dimensionality
of the parameter space to be explored is large, with possibly of order
twenty parameters to be considered. This combination of a limited
number of simulation runs and a rather large number of cosmological
and modeling parameters to be constrained demands a thoughtful design
strategy for deciding on the parameter settings at which to run the
simulations.

 \begin{figure}[b]
  \centering
  \includegraphics[width=3.7in,angle=0] {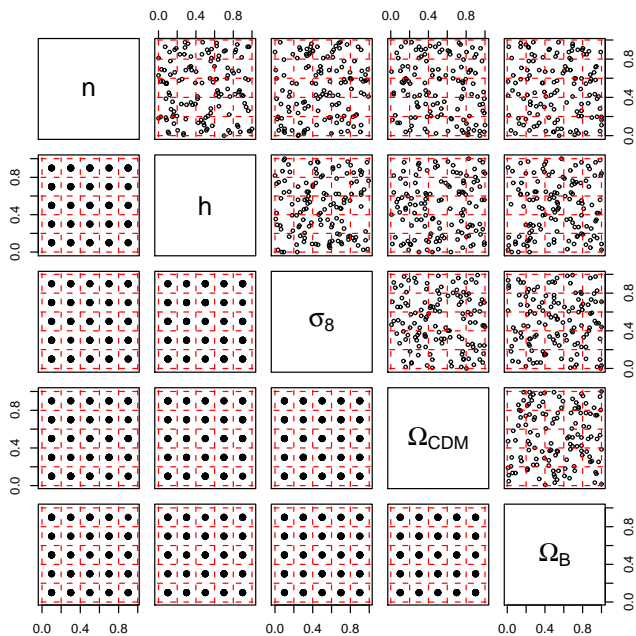}
  \caption{
    \label{fig:oalhs128} Lower triangle of plots: 2-dimensional
    projections of a $m=128$ point, 5-level, OA design. Upper
    triangle: An OA-based LH design obtained by spreading out the 5
    level OA design so that each 1-dimensional projection gives an
    equally spaced set of points along [0,1].}
\end{figure}

The simulation design refers to a sequence of simulation runs carried
out at $m$ input settings which vary over predefined ranges for each
of the $p_\theta$ input parameters: 
\begin{equation}
\label{eq:design}
\begin{pmatrix}
 \theta_{1}^* \\
 \vdots \\
 \theta_{m}^*
\end{pmatrix}
 =
\begin{pmatrix}

\theta_{11}^* & \cdots & \theta_{1p_\theta}^* \\
\vdots        & \vdots & \vdots \\
\theta_{m1}^* & \cdots & \theta_{mp_\theta}^*
\end{pmatrix}.
\end{equation}
We use $\theta^*$ to differentiate the design input settings from the
true value of the parameter vector $\theta$ which is to be estimated.

The design of computer experiments, as simulations are often referred
to in the statistical literature, has received considerable attention
recently in the statistics community, spurred on largely by the
increasing use of complex simulation models to augment understanding
gained from experiments or observations (see Ref.~\cite{sant04},
Chs.~5-6, for a recent survey of the area). The goal in our
application is to use a sequence of simulation runs to build a
GP-based emulator for the expensive simulation code with the aim of
predicting code output at untried parameter input settings. A GP model
typically interpolates the output of the ``training'' simulations
obtained from the experimental design, and gives predictions that vary
smoothly with changes in the input parameters.  Because the GP model
exploits the smoothness in the simulation response (as a function of
the input parameters), space-filling Latin hypercube (LH) designs have
proven to be well suited to the purpose of building GP-based
emulators~{\cite{welch89, ill91}. In particular, we have used orthogonal 
array (OA)-based LH designs~\cite{tang} as well as symmetric 
LH designs~\cite{ye}. Fig.~\ref{fig:oalhs128} shows the $m=128$ point 
design over the $p_\theta=5$ dimensions used in this analysis. This 
design was constructed by perturbing a 5-level orthogonal array design 
so that each 1-dimensional projection gives an equally spaced set of 
points along the standardized parameter range [0,1].

The actual parameter ranges used for the $m=128$ simulations are
\begin{eqnarray}\label{params}
0.8\le &n& \le 1.4,\nonumber\\ 
0.5\le &h& \le 1.1, \nonumber\\ 
0.6 \le &\sigma_8& \le 1.6,\nonumber\\ 
0.05\le &\Omega_{\rm CDM}&\le 0.6,\nonumber\\
0.02\le & \Omega_{\rm b}&\le 0.12.  
\end{eqnarray}
These ranges are standardized to $[0,1]^5$ by shifting and scaling
each interval. (The resulting simulations are produced by joining
spectra obtained from CMBFAST and MC$^2$ as described in Section
\ref{sec:data} and as shown in Fig.\ref{plotone}.)

Such space filling LH designs are well-suited to the strengths of a GP
model which can be fit to an arbitrary set of design points.  In
contrast, more standard interpolation schemes (e.g.,
Ref.~\cite{press99}, Ch.~3.6) typically require a grid-based design
for interpolation.  Grid-based designs are very inefficient since
expanding even a sparse grid over a $p_ \theta$-dimensional space will
be computationally prohibitive. For our example, where $p_\theta = 5$,
even a $[0, \frac{1}{2}, 1]^5$ grid requires 243 simulations.  The
grid-based approach, while simpler, also gives very poor coverage of
low-dimensional projections.  In our example, most of the simulator
activity is explained by two parameters, $\sigma_8$ and $\Omega_{\rm
  CDM}$, for which the grid-based design assigns only 9 unique values.

\subsection{Emulating Simulator Output}
\label{sec:emulator}

Our analysis requires the development of a probability model to
describe the simulator output at untried settings $\theta$. To do
this, we use the simulator outputs to construct a GP model that
``emulates'' the simulator at arbitrary input settings over the
(standardized) design space $[0,1]^{p_\theta}$. To construct this
emulator, we model the simulation output using a $p_\eta$-dimensional
basis representation:
\begin{equation}
 \eta(k;\theta) = \sum_{i=1}^{p_\eta} \phi_i(k) w_i(\theta) + \epsilon,\;
 \theta \in [0,1]^{p_\theta},
\label{eq:etabasis}
\end{equation}
where $\{\phi_1(k),\ldots,\phi_{p_\eta}(k)\}$ is a collection of orthogonal,
$n_\eta$-dimensional basis vectors, the $w_i(\theta)$'s are GPs over
the input space, and $\epsilon$ is an $n_\eta$-dimensional error term.
This type of formulation reduces the problem of building an emulator
that maps $[0,1]^{p_\theta}$ to $R^{n_\eta}$ to building $p_\eta$
independent, univariate GP models for each $w_i(\theta)$. The details
of this model specification are given below.

Output from each of the $m$ simulation runs prescribed by the design
results in $n_\eta$-dimensional vectors, which we denote by
$\eta_1,\ldots,\eta_m$. Since the simulations rarely give incomplete
output, the simulation output can often be efficiently represented via
principal components~\cite{RamSil97}.  We first standardize the
simulations by centering about the mean of raw simulation output
vectors: $\frac{1}{m} \sum_{j=1}^m \eta_j$. We then scale the output
by a single value so that its variance is 1.  This standardization
simplifies some of the prior specifications in our models.  We also
note that, depending on the application, an alternative
standardization may be preferred.  Whatever the choice of the
standardization, the same standardization is also applied to the
experimental data.

We define $y_{\rm sims}$ to be the $n_\eta \times m$ matrix obtained by
column-binding the (standardized) output vectors from the simulations
\begin{equation}
 y_{\rm sims} = [\eta_1;\cdots;\eta_m].
\end{equation}
Typically, the size of a given simulation output $n_\eta$ is much
larger than the number of simulations carried out, $m$.  We apply the
singular value decomposition (SVD) to the simulation output matrix
$y_{\rm sims}$ giving
\begin{equation}
 y_{\rm sims} = UDV^T,
\end{equation}
where $U$ is a $n_\eta \times m$ orthogonal matrix, $D$ is a diagonal
$m \times m$ matrix holding the singular values, and $V$ is a $m
\times m$ orthonormal matrix. To construct a $p_\eta$-dimensional
representation of the simulation output, we define the principal
component (PC) basis matrix $\Phi_\eta$ to be the first $p_\eta$ columns
of $[\frac{1}{\sqrt{m}}UD]$.  The resulting principal component
loadings or weights are then given by $[\sqrt{m}V]$, whose columns
have variance 1.

For representing the mass power spectrum we found that it is adequate
to take $p_\eta = 5$ so that $\Phi_\eta =
[\phi_1;\phi_2;\phi_3;\phi_4;\phi_5]$; the basis functions $\phi_1$,
$\phi_2$, $\phi_3$, $\phi_4$ and $\phi_5$ are shown in
Fig.~\ref{fig:mpseof}.
\begin{figure}[bh]
\centerline{
 \includegraphics[totalheight=3.7in,angle=-90,clip=]{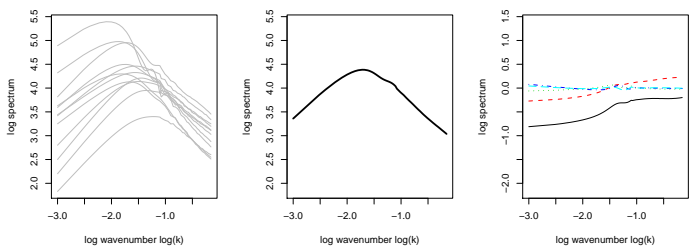}}
\caption{Simulations (left), mean (center), and the first five 
principal component bases (right) derived from the simulation output
for the power spectrum.
\label{fig:mpseof}}
\end{figure}
Note that the $\phi_i$'s are functions of the logarithm of the wave
number, $k$.

We use the basis representation of Eqn.~(\ref{eq:etabasis}) to model
the $n_\eta$-dimensional simulator output over the input space. Each
PC weight $w_i(\theta)$, $i=1,\dots,p_\eta$, is then modeled as a
mean-zero GP
\begin{equation}
w_i(\theta) \sim \mbox{GP}(0,\lambda^{-1}_{wi} R(\theta,\theta';\rho_{wi})
), 
\label{eq:gpw}
\end{equation}
where $\lambda_{wi}$ is the marginal precision of the process and
the correlation function is given by
\begin{equation}
R(\theta,\theta';\rho_{wi}) = \prod_{l=1}^{p_\theta} 
\rho_{wil}^{4(\theta_{l} - \theta'_{l})^2}.
\label{eq:wcor}
\end{equation}
This is the Gaussian covariance function, which gives very smooth
realizations, and has been used previously in Refs.~\cite{KOH, SWMW}
to model simulation output.  An advantage of this product form is that
only a single additional parameter is required per additional input
dimension, while the fitted GP response still allows for rather
general interactions between inputs. We use this Gaussian form for the
covariance function because the simulators we work with tend to
respond very smoothly to changes in the inputs. Depending on the
nature of the sensitivity of simulation output to input changes, one
may wish to alter this covariance specification to allow for rougher
realizations. The parameter $\rho_{wil}$ controls the spatial range
for the $l$th input dimension of the process $w_i(\theta)$. Under this
parameterization, $\rho_{wil}$ gives the correlation between
$w_i(\theta)$ and $w_i(\theta')$ when the input conditions $\theta$
and $\theta'$ are identical, except for a difference of 0.5 in the
$l$th component.  Note that this interpretation makes use of the
standardization of the input space to $[0,1]^{p_\theta}$.

Restricting to the $m$ input design settings given in
Eqn.~(\ref{eq:design}), we define the $m$-vector $w_i$ to be the
restriction of the process $w_i(\cdot)$ to the input design settings
\begin{equation}
 w_i = (w_i(\theta^*_{1}),\ldots,
        w_i(\theta^*_{m}))^T,\;\;
 i=1,\ldots,p_\eta.
\end{equation}
In addition we define $R(\theta^*;\rho_{wi})$ to be the $m \times m$
correlation matrix resulting from applying Eqn.~(\ref{eq:wcor}) to
each pair of input settings in the design.  The $p_\theta$-vector
$\rho_{wi}$ gives the correlation distances for each of the input
dimensions.

At the $m$ simulation input settings, the $m p_\eta$-vector
$w=(w_1^T,\ldots,w_{p_\eta}^T)^T$ then has prior distribution
\begin{equation}
  \begin{pmatrix} w_1 \cr \vdots \cr w_{p_\eta} \end{pmatrix}
  \sim
  N\left( \begin{pmatrix} 0 \cr \vdots \cr 0 \end{pmatrix},
  \begin{pmatrix}
     \Lambda_{w_1}
                                                  & 0 & 0 \cr
     0 & \ddots & 0 \cr
     0 & 0 & \Lambda_{w_{p_\eta}}
  \end{pmatrix}
  \right),
\label{eq:wprior1}
\end{equation}
\[
\Lambda_{w_i}\equiv\lambda^{-1}_{w_i}R(\theta^*;\rho_{w_1}),
\]
which is controlled by $p_\eta$ precision parameters held in
$\lambda_{w}$ and $p_\eta \cdot p_\theta$ spatial correlation
parameters held in $\rho_w$.

The centering of the simulation output makes the choice of zero mean
prior appropriate.  The prior above can be written more compactly as
\begin{equation*}
 w \sim N(0,\Sigma_w),
\end{equation*}
where $\Sigma_w$, controlled by parameter vectors $\lambda_w$ and
$\rho_{w}$, is given in Eqn.~(\ref{eq:wprior1}).

We specify independent $\Gamma(a_w,b_w)$ priors for each $\lambda_{wi}$ and
independent beta$(a_{\rho_w},b_{\rho_w})$ priors for the $\rho_{wil}$'s:
\begin{eqnarray*}
\pi(\lambda_{wi}) & \propto & \lambda_{wi}^{a_w-1} e^{-b_w \lambda_{wi}},
   \;\; i=1,\ldots,p_\eta, \\
\pi(\rho_{wil}) & \propto & \rho_{wil}^{a_{\rho_w}-1} (1-\rho_{wil})^{b_{\rho_w}-1},
   \;\; i=1,\ldots,p_\eta, \\ 
&&\hspace{3.8cm}l=1,\ldots, p_\theta.
\end{eqnarray*}
We expect the marginal variance for each $w_i(\cdot)$ process to be
close to unity due to the standardization of the simulator output.
For this reason we specify that $a_w = b_w= 5$. Thus $\lambda_{wi}$
has a prior mean of 1, and a prior standard deviation of $0.45$.  In
addition, this informative prior helps stabilize the resulting
posterior distribution for the correlation parameters which can be
traded off with the marginal precision parameter~\cite{kern}.
%(Sanso and Stroud,200X).

Because we expect only a subset of the inputs to influence the
simulator response, our prior for the correlation parameters reflects
this expectation of ``effect sparsity''.  Under the parameterization
in Eqn.~(\ref{eq:wcor}), input $l$ is inactive for PC $i$ if
$\rho_{wil}=1$.  Choosing $a_{{\rho_w}}=1$ and $0 < b_{{\rho_w}} < 1$
will give a density with substantial prior mass near 1.  We take
$b_{\rho_w} = 0.1$, which makes Pr$(\rho_{wil} < 0.98) \approx 1/3$
{\em a priori}.  In general, the selection of these hyperparameters
should depend on how many of the $p_\theta$ inputs are expected to be
active.

If we take the error vector in the basis representation of
Eqn.~(\ref{eq:etabasis}) to be i.i.d. (independent and identically
distributed) normal, we can then develop the sampling model, or
likelihood, for the simulator output. We define the $n_\eta m$-vector
$\eta$ to be the concatenation of all $m$ simulation output vectors
\begin{equation}
  \eta = \mbox{vec}(y_{\rm sims})
   =   \mbox{vec}([\eta(\theta_{1}^*);\cdots; \eta(\theta_{m}^*)]).
\end{equation}
Given precision $\lambda_\eta$ of the errors, the likelihood is
then
\begin{equation}
L(\eta|w,\lambda_\eta) \propto
 \lambda_\eta^{\frac{mn_\eta}{2}} \exp \left\{ -\half \lambda_\eta
 (\eta - \Phi w)^T (\eta - \Phi w) \right\}, \;\;
 % \lambda_\eta \sim \Gamma(a_\eta,b_\eta)
\end{equation}
where the $n_\eta \times m p_\eta$  matrix $\Phi$ is given by
\begin{equation}
 \Phi = [I_m \otimes \phi_1; \cdots ; I_m \otimes \phi_{p_\eta}],
\end{equation}
and the $\phi_i$'s are the $p_\eta$ basis vectors previously computed
via SVD. A $\Gamma(a_\eta,b_\eta)$ prior is specified for the error
precision $\lambda_\eta$.  This parameter controls the size of the
errors between the actual simulations and the basis representation of
the simulations.  We expect the data to be very informative about
$\lambda_\eta$, so we choose $a_\eta = 1$ and $b_\eta = .0001$, which
gives little prior information regarding $\lambda_\eta$.

Since the likelihood factors are as shown below
\begin{eqnarray*}
&&L(\eta|w,\lambda_\eta) \propto\\
&& \lambda_\eta^{\frac{mp_\eta}{2}} \exp \left\{ -\half \lambda_{\eta}
 (w - \hat{w})^T (\Phi^T\Phi) (w-\hat{w}) \right\} \times \\
&& \lambda_\eta^{\frac{m(n_\eta-p_\eta)}{2}} \exp \left\{ -\half \lambda_{\eta}
 \eta^T (I-\Phi(\Phi^T\Phi)^{-1}\Phi^T) \eta \right\},
\end{eqnarray*}
the formulation can be equivalently represented with a dimension-reduced
likelihood and a modified $\Gamma(a'_\eta,b'_\eta)$ prior for $\lambda_\eta$:
\begin{eqnarray}
\label{eq:likeeta}
&&L(\hat{w}|w,\lambda_\eta) \propto \nonumber\\
&& \lambda_\eta^{\frac{mp_\eta}{2}} \exp \left\{ -\half \lambda_\eta
 (\hat{w}-w)^T (\Phi^T\Phi
) (\hat{w}-w) \right\},\nonumber\\
\end{eqnarray}
where
\begin{eqnarray}
\nonumber
a'_\eta &=& a_\eta+\frac{m(n_\eta-p_\eta)}{2},\\
\nonumber
b'_\eta &=& b_\eta + \half \eta^T (I-\Phi(\Phi^T\Phi)^{-1}\Phi^T) 
\eta,\mbox{ and}\\
\label{eq:what}
\hat{w} &=& (\Phi^T\Phi)^{-1}\Phi^T \eta.
\end{eqnarray}
Thus, the normal likelihood for $\eta$ with the gamma prior for
$\lambda_\eta$
\begin{equation*}
 \eta|w,\lambda_\eta \sim N(\Phi w,\lambda_\eta^{-1} I_{n_\eta}),\,
 \lambda_\eta \sim \Gamma(a_\eta,b_\eta)
\end{equation*}
is mathematically equivalent to a normal model for
$\hat{w}$ with an altered gamma-prior model for $\lambda_\eta$
\begin{equation*}
 \hat{w}|w,\lambda_\eta \sim
N(w,(\lambda_\eta \Phi^T\Phi)^{-1}),\,
 \lambda_\eta \sim \Gamma(a'_\eta,b'_\eta)
\end{equation*}
since
\begin{equation}
 \label{eq:ngequiv}
 L(\eta|w,\lambda_\eta) \times \pi(\lambda_\eta;a_\eta,b_\eta) \propto
 L(\hat{w}|w,\lambda_\eta) \times \pi(\lambda_\eta;a'_\eta,b'_\eta).
\end{equation}

The likelihood depends on the simulations only through
the computed PC weights $\hat{w}$.  After integrating out
$w$, the posterior distribution becomes
\begin{eqnarray}
\label{eq:postw}
\lefteqn{ \pi(\lambda_\eta,\lambda_w,\rho_w| \hat{w})  \propto }\nonumber\\
  & & \left| (\lambda_\eta \Phi^T\Phi)^{-1} + \Sigma_w \right|^{-\frac{1}{2}}\times\\
  && \exp\{ -\half \hat{w}^T ([\lambda_\eta \Phi^T\Phi]^{-1} 
+ \Sigma_w)^{-1} \hat{w} \}
  \times \\
\nonumber
  & &
  \lambda_\eta^{a'_\eta-1} e^{-b'_\eta \lambda_\eta}
  \prod_{i=1}^{p_\eta} \lambda_{wi}^{a_w-1} e^{-b_w \lambda_{wi}}
  % \\
%\nonumber
 % & &
  \prod_{i=1}^{p_\eta}
       %\left\{
       %\prod_{j=1}^{p_x} (1-\rho_{wij})^{b_\rho-1} \;
       \prod_{j=1}^{p_\theta} (1-\rho_{wij})^{b_\rho-1}.
       %\right\}.
\end{eqnarray}
This posterior distribution is a milepost on the way to the complete
formulation, which also incorporates experimental data.  However, it
is well worth further considering this intermediate posterior
distribution for the simulator response.  It can be explored via MCMC
using standard Metropolis updates and we can view a number of
posterior quantities to illuminate features of the simulator. For
example, Fig.~\ref{fig:rhobox} shows boxplots of the posterior
distributions for the components of $\rho_w$.  From this figure it is
apparent that PCs 1 and 2 are most influenced by $\sigma_8$ and
$\Omega_{\rm CDM}$.
\begin{figure}[ht]
\centerline{
 \includegraphics[width=3.2in,totalheight=7cm,angle=0,clip=]{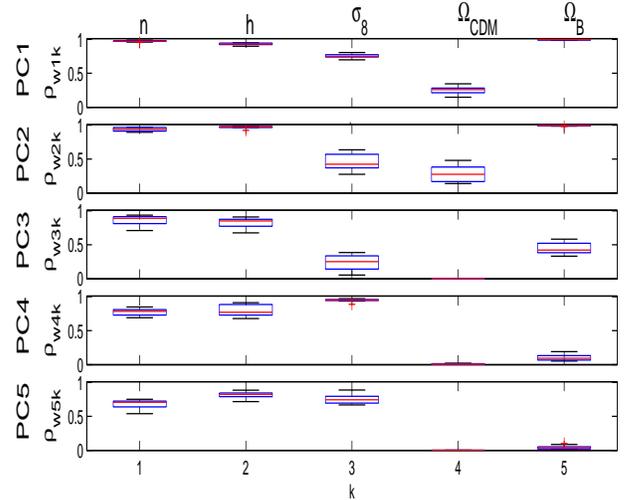}
}
\caption{ Boxplots of posterior samples for each
 $\rho_{wil}$ for the mass power spectrum. 
\label{fig:rhobox}}
\end{figure}

Given the posterior realizations from Eqn.~(\ref{eq:postw}), one can
generate realizations from the process $\eta(\theta)$ at any input
setting $\theta^\star$.  Since
\begin{equation}
\eta(k;\theta^\star) =
  \sum_{i=1}^{p_\eta} \phi_i(k) w_i(\theta^\star),
\end{equation}
realizations from the $w_i(\theta^\star)$ processes need to be drawn
given the MCMC output. For a given draw
$(\lambda_\eta,\lambda_w,\rho_w)$ a draw of 
$w^\star = (w_1(\theta^\star),\ldots,
            w_{p_\eta}(\theta^\star))^T$
can be produced by making use of the fact
\begin{eqnarray}
&& \begin{pmatrix} \hat{w} \cr w^\star \end{pmatrix}
 \sim \nonumber\\
&& N \left( \begin{pmatrix} 0 \cr 0 \end{pmatrix},
   \left[ \begin{pmatrix} (\lambda_\eta \Phi^T\Phi)^{-1} & 0 \cr 0 &
       0 \end{pmatrix} 
   + \Sigma_{w,w^\star}(\lambda_w,\rho_w) \right] \right),\nonumber\\
\end{eqnarray}
where $\Sigma_{w,w^\star}$ is obtained by applying the prior
covariance rule to the augmented input settings that include the
original design and the new input setting $(\theta^\star)$.  [Recall
that $\hat{w}$ is defined in Eqn.~(\ref{eq:what}).]  Application of
the conditional normal rules then gives
\begin{equation}
\label{eq:emulatorpred}
  w^\star|\hat{w} \sim N(V_{21}V_{11}^{-1}\hat{w},
                 V_{22}-V_{21}V_{11}^{-1}V_{12}),
\end{equation}
where
\begin{eqnarray}
  V &=&
  \begin{pmatrix} V_{11} & V_{12} \cr V_{21} &
    V_{22} \end{pmatrix}\nonumber\\
 &=& \left[ \begin{pmatrix} (\lambda_\eta \Phi^T\Phi)^{-1} & 0 \cr 0 & 0 \end{pmatrix}
   + \Sigma_{w,w^\star}(\lambda_w,\rho_w) \right]
\end{eqnarray}
is a function of the parameters produced by the MCMC output.  Hence,
for each posterior realization of $(\lambda_\eta, \lambda_w, \rho_w)$,
a realization of $w^\star$ can be produced.  The above recipe easily
generalizes to give predictions over many input settings at once.

Fig.~\ref{fig:sens03} shows posterior means for the simulator
response $\eta$ where each of the inputs is varied over its prior
range of $[0,1]$ while the other four inputs are held at their nominal
setting of 0.5.
\begin{figure}[ht]
\centerline{
 \includegraphics[width=3.5in,height=1.2in,angle=0,clip=]{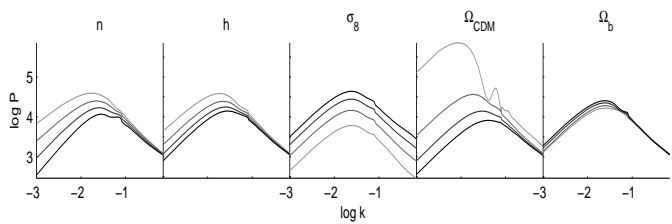}
}
\caption{
  Changes to the posterior mean simulator predictions obtained by
  varying one input, while holding others at their central values,
  i.e. at the midpoint of their range. The light to dark lines
  correspond to the smallest parameter setting to the biggest, for
  each parameter.  \label{fig:sens03} }
\end{figure}
The posterior mean response conveys an idea of how the different
parameters affect the highly multivariate simulation output.  Other
marginal functionals of the simulation response can also be calculated
such as sensitivity indicies or estimates of the Sobol
decomposition~\cite{SWMW,OOH}.

Note that a simplified emulator can be constructed by taking point
estimates for $(\lambda_\eta, \lambda_w, \rho_w)$ (posterior mean, or
posterior medians) and then defining the emulator to be the mean in
Eqn.~(\ref{eq:emulatorpred}).

\subsubsection{Emulator Test and Convergence}

How do we know that the constructed emulator is accurate and how many
model simulations are needed to obtain the desired emulation accuracy?
The answers to these questions depend very much on the smoothness of
the function we wish to emulate. If the function is smooth
and almost featureless, one anticipates that only a small number of
simulations should suffice to yield an accceptable emulator. The
presence of features and noise in the function to be emulated will
clearly require many more simulation outputs; it may well be that
beyond some point the optimal emulation strategy is no longer based on
the GP model. The number of simulations required to build an accurate
emulator depends strongly on the number of active parameters, i.e.,
parameters that change the output considerably when varied.

To test the accuracy of a proposed emulator, so-called hold-out tests
are very useful. The basic idea is simple: A small subset of of the
simulations is set aside and the emulator built on the remaining
simulations. Then the new emulator is evaluated at the parameter
settings of the held-out simulations. By comparing the emulator and
simulation results, the accuracy of the emulator can be estimated. An
example of this approach is provided in Ref.~\cite{HHNH} where we
perform hold-out tests by building the emulator on a subset of 125 out
of the total of 128 simulations. Then we test the the emulator on the
remaining three simulations by running it at the exact parameter
settings of the three holdouts. In this way we test (in turn) the
emulator on all 128 simulations.

In the current paper, we present a slightly modified strategy. In
addition to the 128 run design, we also employ an independent 64 run
design as a reference to investigate the accuracy of the emulator. The
present approach has the advantage that the emulator under test is
built on the full set of simulations. This might not be too important
if the design is already large, but if the design is restricted to a
small number of runs, every run is important in determining the
accuracy of the emulator, especially near the boundaries of the
sampling domain.

In the left panel of Fig.~\ref{fig:resid} we show the results for the
predictions of the 128 run emulator on the additional 64 runs.
Overall, the emulator performance is very satisfactory. We display the
residuals of the emulator prediction compared to the simulation runs
on the native scale. The dark gray band contains the middle 50\% of
the residuals, the light gray band the middle 90\%. The overall
accuracy of our emulator over a wide range in wave number is better
than $\sim 5$\%. Only on the edges of the parameter ranges
investigated is the fidelity slightly worse. Note that the emulation
accuracy is strongly dependent on the size of the allowed parameter
range, and improves significantly as this range is shrunk. We will
come back to this point later below.

Next we investigate how the accuracy of the emulator depends on the
number of the underlying simulations. We create a design for 32
simulations using an OA-LH sampling design and test it on the same
reference 64 design run we already used for testing the 128 run
emulator. In Fig.~\ref{fig:resid}, right panel, we show the same
statistics as for the 128 run emulator. The overall quality of the
emulator is still very good, at the 10\% level. Compared with the
larger-design emulator, however, the predictions in the medium $k$
range show a higher level of deviation from the reference values.

\begin{figure}
\begin{center}
\includegraphics[totalheight=85mm,width=60mm,angle=-90]{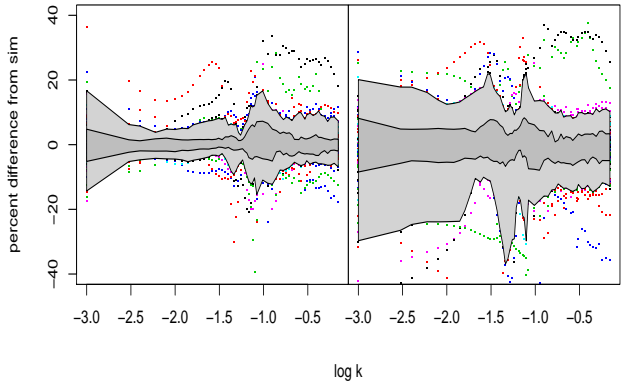}
\end{center}
\caption{Emulator predictions for a 64 run design. Left: emulator
  based on 128 runs, right: emulator based on 32 runs. The central
  gray region contains the middle 50\% of the residuals, the wider
  light gray region, the middle 90\%. The outliers are shown as
  dots. The improvement of the emulator accuracy with more training
  runs is evident, especially in the medium $k$-range.}
\label{fig:resid}
\end{figure}

\subsection{Full Statistical Formulation}
\label{sec:full}

Given the model specifications for the simulator $\eta(\theta)$, we
can now consider the sampling model for the experimentally observed
data.  The data are contained in an $n$-vector $y$.  For the synthetic
mass power spectrum application, $n=28$, corresponding to the
different wave numbers as shown in Fig.~\ref{plotone}.  As stated in
Section~\ref{sec:stat}, the data are modeled as a noisy version of the
simulated spectrum $\eta(k;\theta)$ run at the true, but unknown,
parameter setting $\theta$, $y=\eta(k;\theta) + \epsilon,$ where the
errors are assumed to be $N(0,\Sigma_y)$.  For notational convenience
we represent $\Sigma_y^{-1}$ as $\lambda_y W_y$, leaving open the
option to estimate a scaling of the error covariance with $\lambda_y$.
Using the basis representation for the simulator this becomes
\begin{eqnarray*}
 y &=& \Phi_y w(\theta) + \epsilon,
\end{eqnarray*}
where $w(\theta)$ is the $p_\eta$-vector
$(w_1(\theta),\ldots,w_{p_\eta}(\theta))^T$.  Because the wave number
support of $y$ is not necessarily contained in the support of the
simulation output, the basis vectors in $\Phi_y$ may have to be
interpolated over wave number from $\Phi_\eta$.  Since the simulation
output over wave number is quite dense, this interpolation is
straightforward.

We specify a $\Gamma(a_y,b_y)$ prior for the precision parameter
$\lambda_y$ resulting in a normal-gamma form for the data model
\begin{equation}
\label{eq:datalike}
 y| w(\theta), \lambda_y \sim
     N( \Phi_y w(\theta),(\lambda_y W_y)^{-1}),\,
 \lambda_y \sim \Gamma(a_y,b_y).
\end{equation}
The observation precision $W_y$ is often fairly well-known in
practice. Hence one may choose to fix $\lambda_y$ at 1, or use an
informative prior that encourages its value to be near one.  In the
mass power spectrum example we fix $\lambda_y$ at 1, though we
continue to use this parameter in the formulation below.

Equivalently, Eqn.~(\ref{eq:datalike}) can be represented in terms
of the basis weights
\begin{equation}
\label{eq:datalikered}
 \hat{w}_y | w(\theta), \lambda_y 
 \sim
 N(w(\theta),(\lambda_y \Phi_y^T W_y \Phi_y)^{-1}),\;
  \lambda_y \sim \Gamma(a'_y,b'_y),
\end{equation}
with
\begin{eqnarray*}
  \hat{w}_y &=& (\Phi_y^T W_y \Phi_y)^{-1} \Phi_y^T W_y y, \\
    a'_y &=& a_y + \half (n -  p_\eta), \mbox{ and}\\
    b'_y &=& b_y + \half (y-\Phi_y\hat{w}_y)^T W_y (y-\Phi_y\hat{w}_y).
    \end{eqnarray*}
This equivalency follows from Eqn.~(\ref{eq:ngequiv}) given in
Sec.~\ref{sec:emulator}. 

The (marginal) distribution for the combined, reduced data obtained
from the experiments and simulations given the covariance parameters,
has the form
\begin{equation}
\label{eq:vuwdist}
  \begin{pmatrix} \hat{w}_y \cr \hat{w} \end{pmatrix} \sim
  \mbox{N}\left( \begin{pmatrix} 0 \cr 0 \end{pmatrix},
     \begin{pmatrix}
       \Lambda_y^{-1} &  0 \cr 
        0 & \Lambda_\eta^{-1}
     \end{pmatrix}
      +
     \begin{pmatrix}
        I_{p_\eta} & \Sigma_{w_y w}     \cr
        \Sigma_{w_y w}^T & \Sigma_{w}
     \end{pmatrix}
   \right),
\end{equation}
where $\Sigma_w$ is defined in (\ref{eq:wprior1}),
\begin{eqnarray*}
 \Lambda_y & = & \lambda_y \Phi_y^T W_y \Phi_y, \\
 \Lambda_\eta & = & \lambda_\eta \Phi^T\Phi, \\
 I_{p_\eta} &=& p_\eta \times p_\eta \mbox{ identity matrix},\\
 \Sigma_{w_y w} &=& \begin{pmatrix}
     \lambda^{-1}_{w1}R(\theta,\theta^*;\rho_{w1}) & 0 & 0 \\
     0 & \ddots & 0 \\
     0 & 0 & \lambda^{-1}_{wp_\eta}R(\theta,\theta^*;\rho_{wp_\eta}) 
     \end{pmatrix}.
 \end{eqnarray*}
 Above, $R(\theta,\theta^*;\rho_{wi})$ denotes the $1 \times m$
 correlation submatrix for the GP modeling the simulator output
 obtained by applying Eqn.~(\ref{eq:wcor}) to the observational
 setting $\theta$ crossed with the $m$ simulator input settings
 $\theta^*_1,\ldots,\theta^*_m$.

\begin{figure}
\centerline{
 \includegraphics[width=3.4in,totalheight=3.1in,angle=0,clip=]{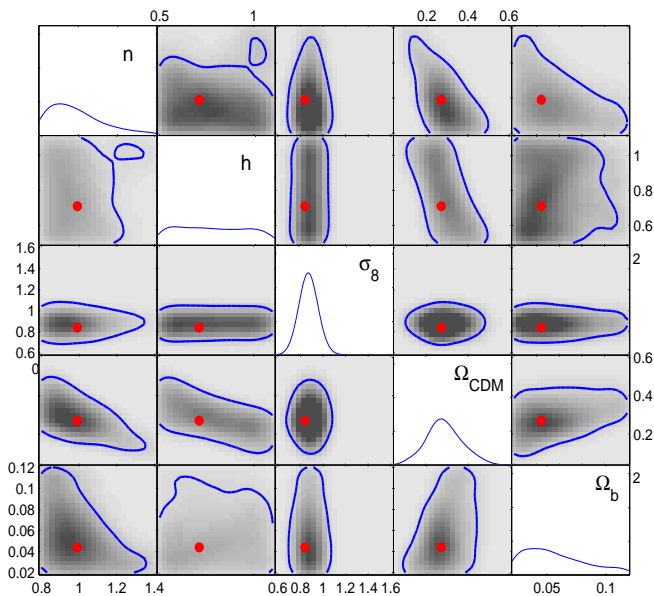}
}
\caption{Estimated posterior distribution of the parameters $\theta =
  (n,h,\sigma_8,\Omega_{\rm CDM},\Omega_{\rm b})$.  The diagonal shows
  the estimated marginal posterior pdf for each parameter; the
  off-diagonal images give estimates of bivariate marginals; the
  contour lines show estimated 90\% hpd regions.  The true values from
  which the data were generated are shown by the red
  dots.  \label{fig:posttheta} }
\end{figure}

\subsubsection{Posterior distribution}
\label{sec:post}

If we take $\hat{z}$ to denote the reduced data
$(\hat{w}^T_y,\hat{w}^T)^T$, and $\Sigma_{\hat{z}}$
to be the covariance matrix given in Eqn.~(\ref{eq:vuwdist}),
the posterior distribution has the form
\begin{eqnarray}
 \label{eq:fullpost}
 \lefteqn{\pi(\lambda_\eta,\lambda_w,\rho_w,\lambda_y,
                    \theta|\hat{z}) \propto }\\
 \nonumber
 &&
 |\Sigma_{\hat{z}}|^{-\frac{1}{2}}
   \exp\left\{ -\half \hat{z}^T \Sigma_{\hat{z}}^{-1} \hat{z} \right\}
 \times \nonumber\\
&&\nonumber \lambda_\eta^{a'_\eta-1} e^{-b'_\eta \lambda_\eta}
 \prod_{i=1}^{p_\eta} \lambda_{wi}^{a_{w}-1} e^{-b_{w} \lambda_{wi}}
 \times \\ \nonumber
 &&
 \prod_{i=1}^{p_\eta} \prod_{l=1}^{p_\theta}
    \rho_{wil}^{a_{\rho_w}-1} (1- \rho_{wil})^{b_{\rho_w}-1}
 \lambda_y^{a'_y-1} e^{-b'_y \lambda_y} I[\theta \in C],
\end{eqnarray}
where $C$ denotes the constraint region for $\theta$, which is
typically a $p_\theta$-dimensional rectangle.  In other applications
$C$ can also incorporate constraints between the components of
$\theta$.

Realizations from the posterior distribution are produced using
standard, single site MCMC. Metropolis updates \cite{metrop} are used
for the components of $\rho_w$ and $\theta$ with a uniform proposal
distribution centered at the current value of the parameter.  The
precision parameters $\lambda_\eta$, $\lambda_w$ and $\lambda_y$ are
sampled using Hastings updates~\cite{hastings}.  Here the proposals
are uniform draws, centered at the current parameter values, with a
width that is proportional to the current parameter value.  In a given
application the candidate proposal width can be tuned for optimal
performance.

The resulting posterior distribution estimate for $\theta$ is shown in
Fig. \ref{fig:posttheta} on the original scale.  It nicely brackets
the true values of $\theta = (0.99,0.71,0.84,0.27,0.044)$ from which
the synthetic data were generated.

\section{Combined CMB and large scale structure analysis}
\label{sec:combined}

So far our analysis has focused on a single observational data set,
but cosmological parameter estimation requires combining a number of
different observational datasets with their (separate) associated
modeling methodologies. We now consider a simple example of this, the
joint analysis of CMB temperature anisotropy data and the mass power
spectrum as sampled by large-scale structure surveys.

The power spectrum of the CMB anistropy as a function of angular scale
or multipole moment $l$ has significantly more structure than the
matter power spectrum and therefore presents a more challenging test
for our framework. In this section we extend our synthetic data set to
include these measurements. We restrict our analysis to the
six-dimensional ``vanilla'' $\Lambda$CDM model described by
\begin{equation}
\theta=(n,h,\sigma_8,\Omega_{\rm CDM},\Omega_{\rm b},\tau).
\end{equation}
Different groups analyzing different data sets, e.g.
Refs.~\cite{spergel,tegmark06}, found that the model specified by
these six parameters consistently fits all currently available
data. Our synthetic data set mimics data from WMAP-III~\cite{spergel}
and the galaxy mass power spectrum from the SDSS~\cite{tegmark}. For
the synthetic dataset, we choose the same cosmology as given in
Eqn.~(\ref{cosm}) and in addition the optical depth:
\begin{equation}\label{tau}
\tau=0.09.
\end{equation}
We allow $\tau$ to vary between 0 and 0.3 in our analysis below. The
analysis of the CMB data was carried out using the CAMB~\cite{camb}
code throughout. We have carefully checked that the transfer functions
generated with CMBFAST for the $N$-body simulations agree very
accurately with the CAMB transfer functions. Our synthetic data set
allows us to ignore real-world corrections such as for the
normalization from the Sunyaev-Zel'dovich effect due to hot gas in
clusters. The WMAP-III analysis showed that such considerations can
become important for precision measurements of cosmological
parameters.

\subsection{Constraints from the Cosmic Microwave Background}

Before we carry out a combined analysis of galaxy survey and CMB data,
we investigate how well our framework does in dealing with the CMB
temperature angular power spectrum (TT). We do not consider CMB
polarization here but there is no obstacle to including it in future
studies.

As done earlier for the matter power spectrum we generate a synthetic
data set for the TT spectrum. For the error bars we assume the same
magnitude as in the WMAP-III analysis. We first run CAMB at the
parameter settings specified in Eqns.~(\ref{cosm}) and (\ref{tau}) and
then move the data points off the base TT spectrum according to a
Gaussian distribution with 1-$\sigma$ confidence level. Following the
matter power spectrum analysis we create a design for 128 runs, this
time for a six-parameter space. In Fig.~\ref{fig:simstt} we show the
analog to Fig.~\ref{plotone} for the TT spectrum: the gray lines show
a subset of the 128 CAMB runs, the dashed line the actual input power
spectrum and the red points the data points which will be used for our
analysis.
 
\begin{figure}[b] \centering
\includegraphics[totalheight=2.8in,angle=0] {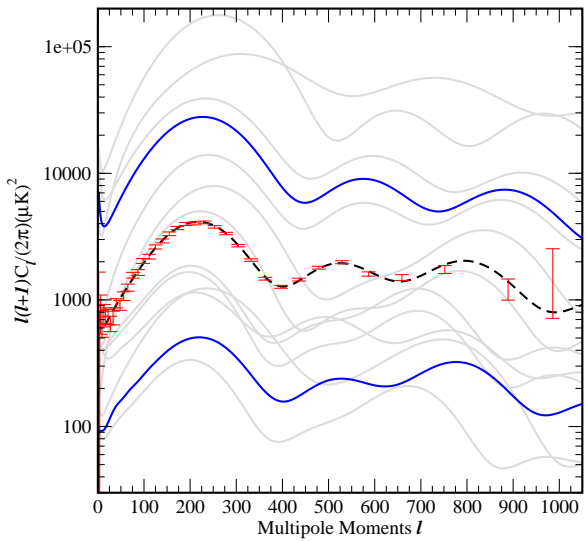} 
\caption{\label{fig:simstt} Subset of 128 simulated TT power spectra
  (gray lines) along with the synthetic observations.  The dashed line
  shows the actual spectrum from which the data were generated.  The
  blue lines show the limits of the restricted parameter ranges from
  Eqns.~(\ref{params_res}).}
\end{figure}

In Fig.~\ref{fig:emulator} we show the analog to Fig.~\ref{fig:resid},
demonstrating that the $C_l$ emulator predicts 90\% of the holdout
runs to better than 10\% and 50\% of the runs to better than 5\%. This
accuracy is impressive considering the dynamic range and complexity of
the $C_l$s (and the broad range over which the cosmological
parameters are varied). However, this complexity requires more PCs for
accurate emulation. We have kept twelve PCs for the $C_l$ analysis
compared to five PCs for the matter power spectrum analysis.  The
bivariate marginal plot summarizing the inference in the cosmological
parameters taking into account the TT spectrum alone is given in
Fig.~\ref{fig:postcls} (compare to Fig.~\ref{fig:posttheta}).

\begin{figure}
  \centering
  \includegraphics[totalheight=2.7in,angle=0]{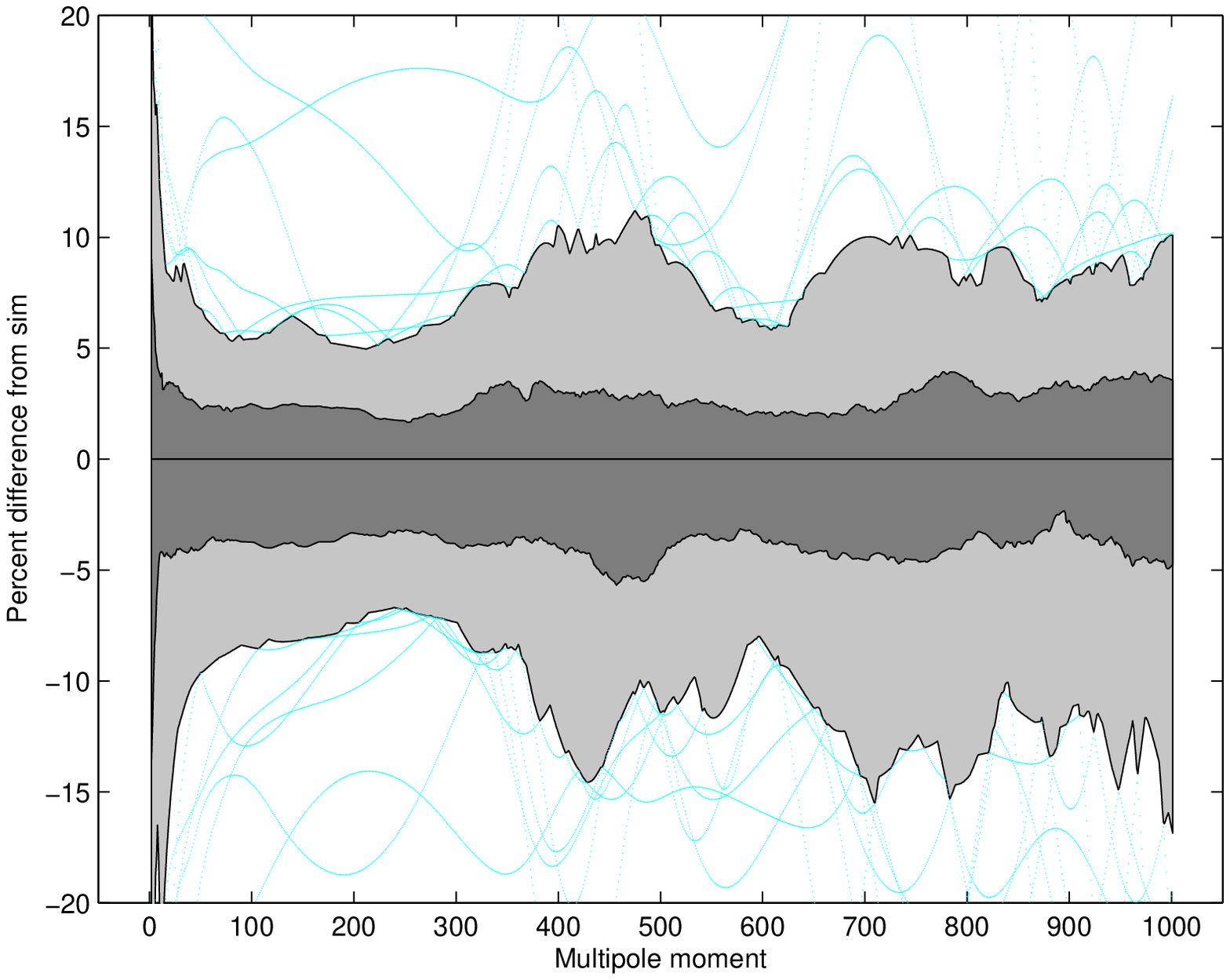}
  \includegraphics[totalheight=2.7in,angle=0] {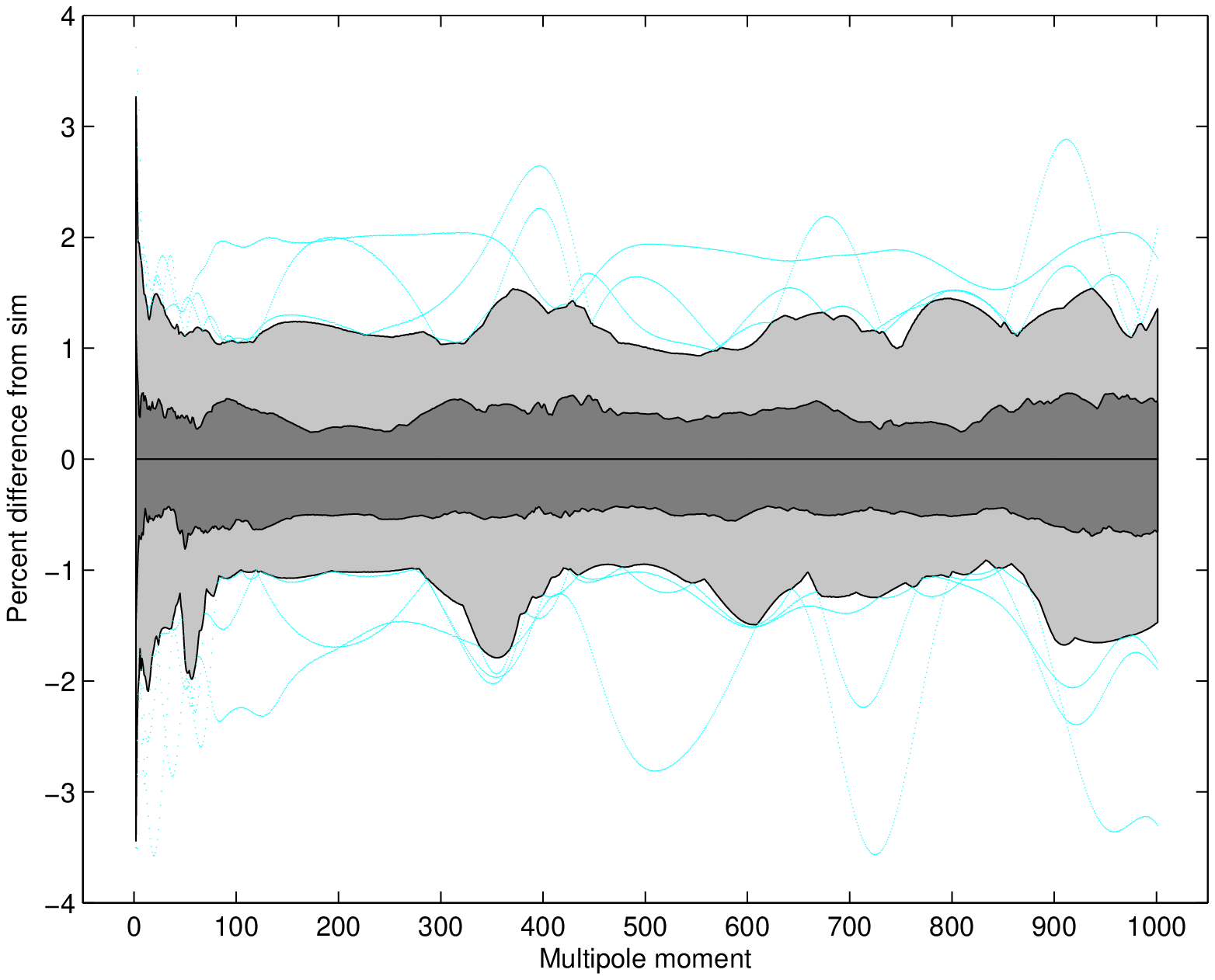}
  \caption{
    \label{fig:emulator} Emulator performance on holdout test.  The
    central gray region contains the middle 50\% of the residuals, the
    wider light gray region, the middle 90\%. The outliers are shown
    as blue curves. Upper plot: Residuals for the emulator built on
    the conservative priors given in Eqns.~(\ref{params}). Lower plot:
    Residuals for the emulator built on 3-$\sigma$ priors around the
    best-fit WMAP-III parameters. The emulator errors are reduced by
    an order of magnitude for the smaller parameter ranges.}
\end{figure}

Other groups who have developed interpolation schemes to predict the
temperature power spectrum~\cite{kap02,Jim04,kos02,fendt06,auld06}
choose much narrower priors than we have in our example. If, as is
typical, we reduce the parameter ranges to 3-$\sigma$ around the
best-fit WMAP data, leading to:
\begin{eqnarray}\label{params_res}
0.85\le &n& \le 1.25,\nonumber\\ 
0.6\le &h& \le 0.9, \nonumber\\ 
0.6 \le &\sigma_8& \le 1.2,\nonumber\\ 
0.06\le &\Omega_{\rm CDM}h^2&\le 0.2,\nonumber\\
0.018\le & \Omega_{\rm b}h^2&\le 0.034,\nonumber\\
0.01\le&\tau&\le 0.55,
\end{eqnarray}
the dynamic range of the $C_l$s is in turn reduced by an order of
magnitude (blue lines in Fig.~\ref{fig:simstt}). This leads to an
improvement of the emulator quality by an order of magnitude to yield
results at sub-percent level accuracy (Fig.~\ref{fig:emulator}, lower
panel). We stress that our parameter ranges in this case are still
larger than what is considered by other groups. Their procedures are
restricted to be valid at 3-$\sigma$ around the best-fit WMAP model,
covering only a small range in parameter space. We compare our method
with the other approaches in more detail in Appendix~\ref{appa}.

\begin{figure}[ht]
  \centering
  \includegraphics[totalheight=2.9in,width=2.9in,angle=0]
  {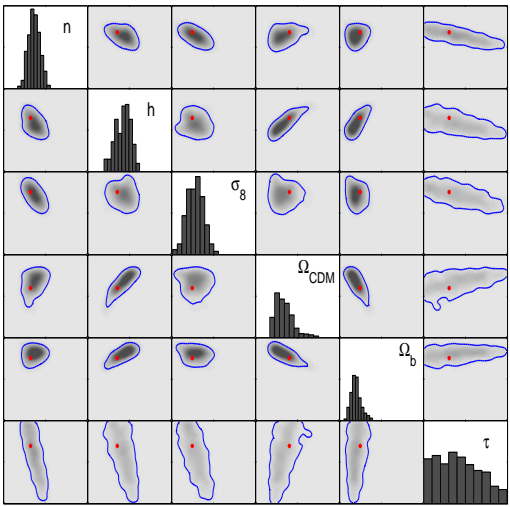} \caption{\label{fig:postcls} Estimated
    posterior distribution following Fig.~\ref{fig:posttheta} for the
    synthetic CMB TT spectrum and including the optical depth $\tau$.}
\end{figure}

\subsection{Combined Constraints}

We now proceed to systematically combine information contained in the
TT and matter power spectra. Given two sets of observed data, $y_1$
and $y_2$ that inform on a common set of parameters, $\theta$, the
posterior density is
\begin{equation} \label{eq:post1}
\pi(\theta|y_1,y_2) \propto L(y_1,y_2|\theta)\cdot\pi(\theta).
\end{equation}
Assuming statistical independence of the two datasets, the likelihood
factors into:
\begin{equation} \label{eq:postproduct}
\pi(\theta|y_1,y_2) \propto L(y_1|\theta)\cdot L(y_2|\theta)\cdot \pi(\theta),
\end{equation}
which is the form we have incorporated into our statistical analysis code.

The payoff from including both sets of observational data is
illustrated in Figs.~\ref{fig:postcomb} and \ref{fig:univdens} and in
Table~\ref{tab1}. The posterior volume of the cosmological parameter
space is significantly reduced by the addition of the matter power
spectrum constraints. There are two main reasons for this volumetric
reduction. First is the expected statistical increase due to the
addition of two independent and consistent pieces of data. Second, and
more interesting, is the influence of posterior correlations among the
cosmological parameters induced by the two datasets. For example, the
degeneracy between $n$ and $\tau$ allows the inclusion of matter power
spectrum data (the simulations of which are independent of $\tau$) to
significantly improve the estimation of $\tau$.

\begin{table*}[t]
  \caption{\label{tab1} Parameter constraints (mean value $\pm$ 1
  standard deviation) for six parameters from CMB TT power spectrum
  analysis alone and with the adddition of large scale structure
  data. The first line gives the true value of the parameters.}
\begin{ruledtabular}
\begin{tabular}{lcccccc}
\hfill & $n$ & $h$ & $\sigma_8$ & $\Omega_{\rm CDM}$ & $\Omega_{\rm b}$ & $\tau$\\
\hline
Truth & 0.99 & 0.71 & 0.84 & 0.27 & 0.044 & 0.09 \\ 
CMB only & 1.0185 $\pm$ 0.0422 & 0.7527 $\pm$ 0.0544 & 0.8824 $\pm$ 0.1004 &
0.2387 $\pm$ 0.0592 & 0.0397$\pm$ 0.006 & 0.1241 $\pm$ 0.0763 \\
CMB + LSS & 1.0090$\pm$ 0.0263 & 0.7335 $\pm$ 0.0371 & 0.8722 $\pm$ 0.0262 &
0.2560 $\pm$ 0.0387 & 0.0413 $\pm$ 0.0044 & 0.0922 $\pm$ 0.0434
\end{tabular}
\end{ruledtabular}
\end{table*}

It is easy to see that including additional datasets into this method
is straightforward.
\begin{figure}
  \centering \includegraphics[totalheight=2.9in,width=2.9in,angle=0]
  {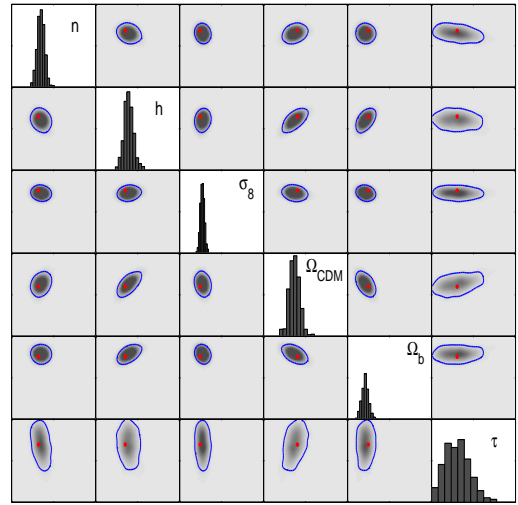} \caption{\label{fig:postcomb} Estimated
    posterior distribution following Fig.~\ref{fig:postcls} with the
    addition of matter power spectrum data.
   }
\end{figure}

\begin{figure}
  \centering
  \includegraphics[totalheight=3.5in,width=2.9in,angle=0]
  {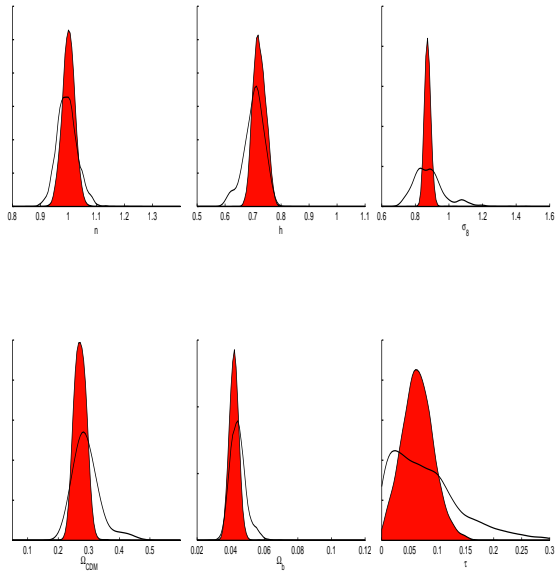} 
  \caption{
    \label{fig:univdens}Posterior univariate marginal density
    estimates for the cosmological parameters taking into account the
    TT only data (white) and both the TT and the matter power spectrum
    data (orange), showing the increased precision obtainable by
    systematically including multiple sources of data.  }
\end{figure}

\section{Discussion and Conclusion}
\label{sec:dis}

We have demonstrated a powerful and general statistical methodology
for performing computer simulation-based inference of cosmological
parameters. To do this, we borrow techniques from a variety of
statistical fields -- including experimental design, spatial
statistics and Kriging, and Bayesian inference -- and apply them in a
highly integrated manner to the problem of constraining computational
models directly from the observed data. Several items are particularly
noteworthy.  First, careful simulation design prevents combinatorial
explosions of simple grid-like designs in highly multivariate
environments. Second, the GP-based emulator design is critical to the
efficient sampling of the posterior probability density and allows us
to evaluate global measures of the simulator sensitivity cheaply and
accurately.  Finally, our method isolates the separate sources of
uncertainty in any particular inference, e.g., the uncertainty due to
imperfect emulation, and folds them into the overall inference
uncertainty. This procedure has the conceptual advantage of coherence
and the practical advantage of guarding against overly optimistic
parameter estimates. It can be easily extended to very large parameter
spaces and to include different data sources such as CMB and large
scale structure measurements.

Depending on the parameter ranges considered, our emulation scheme can
perform at the sub-percent level accuracy, thereby, at least in
principle, satisfying a fundamental requirement for next-generation
cosmological analysis tools. It is especially suited to -- and
designed for -- problems where the underlying simulations are very
costly and only a limited number can be performed. While clearly of
more general utility, our framework is targeted to analysis of
upcoming large-scale structure based probes of dark energy, such as
weak lensing and baryon acoustic oscillations. Future directions and
the usage for our framework in cosmology are manifold. We plan to
extend the set of cosmological parameters under consideration to
include the dark energy equation of state parameter $w$ and to
publicly release an emulator scheme based on very accurate,
high-resolution dark matter simulations. The framework can also be
used to fit supernova light-curves or determine photometric galaxy
redshifts based on training sets. Work in this direction is already in
progress.

\acknowledgments We thank Kevork Abazajian, Derek Bingham, Daniel
Eisenstein, Josh Frieman, Lam Hui, Adam Lidz, and Martin White for
useful discussions and encouragement. We would like to thank the ``Hot
Rocks Cafe'' for providing a pleasant and inspiring collaboration
space. A special acknowledgment is due to supercomputing time awarded
to us under the LANL Institutional Computing Initiative. Finally, we
would like to thank Antony Lewis for help with CAMB, Chad Fendt for
help with Pico, and Licia Verde for clarification of WMAP-III data set
issues. This research is supported by the DOE under contract
W-7405-ENG-36.

\begin{appendix}
\section{Comparison with other methods}
\label{appa}

In this Appendix we compare the performance of our emulator for the TT
power spectrum with other recently introduced interpolation schemes.
An early attempt is based on the idea of splitting the TT power
spectrum into low-$l$ ($l\le 100$) and high-$l$ regions ($l\ge 100$)
and to use analytic fits to express the $C_l$s~\cite{teg00}. This
method is accurate at the 10\% level and forms the basis of the
approach underlying DASh (Davis Anisotropy Shortcut) developed in
Ref.~\cite{kap02}. DASh relies on rapid analytic and semi-analytic
approximations and leads to good accuracy (at the 2\% level on
average) and performance. Another interpolation scheme,
CMBwarp~\cite{Jim04}, is based on introducing a new set of
cosmological parameters~(for details on these parameter choices see
Ref.~\cite{kos02}) which better reflect the underlying physics of the
CMB. The degeneracy structure in the new parameter space is much
simpler and therefore helps with MCMC convergence. The new parameter
set has almost-linear influence on the TT power spectrum (i.e., the
spectrum moves mainly vertically or horizontally with parameter
changes). It is therefore relatively easy to find a polynomial fit
around a fiducial model that is reliable for different
parameters. CMBwarp is faster than DASh at the same level of accuracy.
Both approaches have two major drawbacks: they only lead to accurate
results in a narrow parameter range around a fiducial model and
incorporation of new parameters is very difficult since new fits have
to be developed. In the case of CMBwarp, any additional parameter has
to be ``orthogonal'' to the others, which might be difficult to
achieve. Very recently, a new interpolation method, Pico, was
introduced in Ref.~\cite{fendt06}. Pico is based on a large number of
training sets (a $\sim 10^4$ run MCMC chain). It allows very accurate
and very fast determination of temperature power spectra around the
best-fit WMAP model. Pico is trained to compute the power spectra
within several log likelihoods around the peak. Therefore, good
performance away from the peak cannot be guaranteed. Integration of
new parameters is possible by generating new training data sets. Very
similar to the Pico approach is CosmoNet~\cite{auld06}. In this
approach, a neural network is trained on 2,000 CAMB runs. The CosmoNet
results are very similar in accuracy and speed to Pico.

\begin{figure}
  \centering \includegraphics[totalheight=2.6in,width=3.2in,angle=0]
  {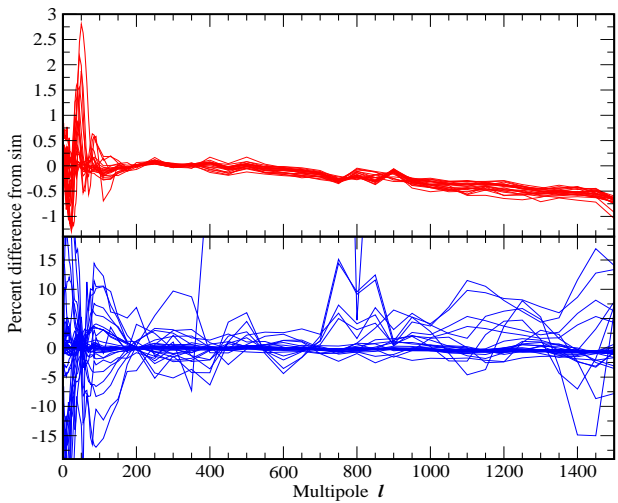} \caption{\label{fig:pico}Residuals for the Pico
  emulator of the TT power spectrum, tested on 64 runs. The upper plot
  in red shows the residuals for runs which are in the 3-$\sigma$
  confidence level around the best-fit WMAP-I model. 22 out of the 64
  runs fulfill this criterion. The residuals are at the 0.5\% accuracy
  level in this case. Note the slight downward trend in all cases. The
  lower plot in blue shows the remaining 42 runs which are in the
  3-$\sigma$ confidence level around the best fit parameters from
  WMAP. In this case the errors are much worse and the emulator is not
  controlled. Again a slight downward trend can be seen in all the
  runs.}
\end{figure}

\begin{figure}
  \centering \includegraphics[totalheight=2.6in,width=3.2in,angle=0]
  {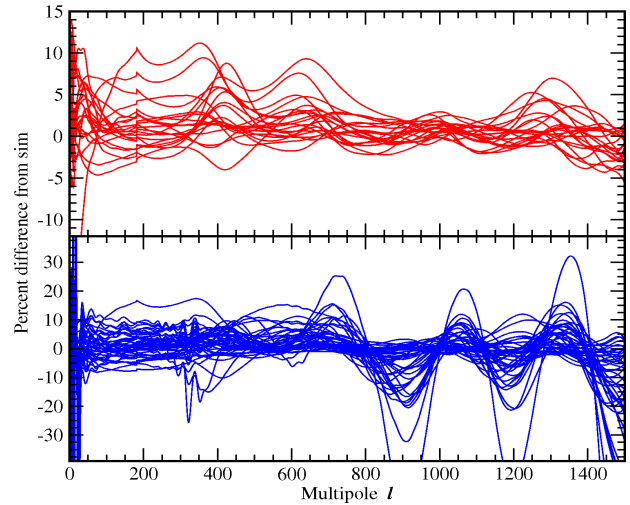} 
  \caption{\label{fig:warp} Results following Fig.~\ref{fig:pico}
    but for the CMBwarp emulator. The errors are overall an order of
    magnitude bigger than for Pico, in agreement with the findings
    of Ref.~\cite{fendt06}.}
\end{figure}

We concentrate our comparison on two publicly available codes: CMBwarp
and Pico. Both codes are designed to yield reliable, accurate results
within WMAP's first-year 3-$\sigma$ confidence region around the
best-fit model. The parameter ranges we have investigated so far are
much broader. Therefore, in order to be able to carry out a meaningful
comparison between the three different approaches, we build a new
emulator based on 128 CAMB runs in the parameter range specified in
Eqn.~(\ref{params_res}). (Note that we now also include $h^2$ in our
variable choice for the dark matter and baryon content of the
Universe). These parameter ranges are within 3-$\sigma$ about the best
fit parameters, which is still a larger range than Pico and CMBwarp
allow. In fact, their allowed ranges are rather narrow and cannot be
cast in the form of a symmetric box around the best fit parameters.

In addition to the 128 training runs we generate 64 reference CAMB
runs for testing our emulator as well as CMBwarp and Pico. In
Fig.~\ref{fig:pico} we show the emulator quality for Pico. Only 22 out
of the 64 simulations lie in the allowed parameter range; we
display the residuals for these runs in the upper part of
Fig.~\ref{fig:pico} in red. The accuracy in this parameter range is
very good, at the 0.5\% level. Somewhat worrisome though is the
existence of a systematic trend in the Pico data: All residuals are
systematically low for large $l$, possibly leading to biases in the
parameter constraints. In the lower plot of Fig.~\ref{fig:pico} we
show the residuals for the remaining 42 simulations in blue, which lie
in the 3-$\sigma$ interval around the best fit parameters but not the
best fit model for WMAP-I. Here the errors are much bigger, by more than
an order of magnitude. This shows that Pico should not be taken out of
the parameter range it was trained for, and is not robust against
extrapolation into larger parameter ranges (as expected for a
polynomial-based fit). Fig.~\ref{fig:warp} shows
the residuals for the CMBwarp runs. As for Pico we have divided them
into runs which are inside the 3-$\sigma$ confidence level around the
best fit model and those outside, but still inside the 3-$\sigma$ confidence
level of the best fit parameters. In general, the performance of CMBwarp
is an order of magnitude worse than the performance of Pico,
confirming the results in Ref.~\cite{fendt06}. For $l>180$, the fit is
slightly modified, leading to a small kink at this value in the
results for the $C_l$s. The performance on all 64 runs of our emulator
scheme based on GP models is shown in Fig.~\ref{fig:emulator}. 50\%
of all runs are predicted with sub-percent accuracy, 90\% with an
accuracy between one and two percent. Not a single prediction is off
by more than 3\%. This demonstrates impressively the stability of our
emulator scheme over large parameter ranges.

\end{appendix}

\end{document}